\documentclass[aps,prl,twocolumn,superscriptaddress,floatfix]{revtex4-1}
\usepackage{hyperref}
\usepackage{bm}
\usepackage{graphicx}
\usepackage[section]{placeins}
\usepackage{nag}
\usepackage{natbib}
\usepackage[geometry]{ifsym} 
\usepackage{color} 

\definecolor{mycyan}{RGB}{0, 255, 255} 
\definecolor{myred}{RGB}{255, 0, 0} 
\definecolor{myblue}{RGB}{0, 0, 255} 

\begin{document}
%
%
\title{Entropy of water and the temperature-induced stiffening of amyloid networks} 
%
%
%
\author{Slav A. Semerdzhiev}
\author{Saskia Lindhoud}
\author{Anja Stefanovic}
\affiliation{Nanobiophysics, MESA+ Institute for Nanotechnology and MIRA Institute for Biomedical Technology and Technical Medicine, University of Twente, P.O. Box 217, 7500 AE Enschede, The Netherlands.\medskip}

\author{Vinod Subramaniam}
\affiliation{Vrije Universiteit Amsterdam, De Boelelaan 1105, 1081 HV Amsterdam, The Netherlands.\medskip}

\author{Paul van der Schoot}
\affiliation{Theory of Polymers and Soft Matter, Eindhoven University of Technology, P.O. Box 513, 5600 MB Eindhoven, The Netherlands \bigskip}

\author{Mireille M.A.E. Claessens*\bigskip}
\affiliation{Nanobiophysics, MESA+ Institute for Nanotechnology and MIRA Institute for Biomedical Technology and Technical Medicine, University of Twente, P.O. Box 217, 7500 AE Enschede, The Netherlands.\medskip}


\date{\today}

\begin{abstract}
In water, networks of semi-flexible fibrils of the protein $\alpha$-synuclein stiffen significantly with increasing temperature. We make plausible that this reversible stiffening is a result of hydrophobic contacts between the fibrils that become more prominent with increasing temperature. The good agreement of our experimentally observed temperature dependence of the storage modulus of the network with a scaling theory linking network elasticity with reversible crosslinking enables us to quantify the endothermic binding enthalpy and an estimate the effective size of hydrophobic patches on the fibril surface.
\end{abstract}


\maketitle 

\section{Introduction}

Water and oil do not mix. Even after vigorous stirring, these two compounds spontaneously separate into distinct liquid phases. This is a macroscopic manifestation of the hydrophobic effect, a phenomenon driven by the microscopic behavior of water in the presence of nonpolar molecules. Ultimately, this is caused by hydrophobic molecules and assemblies thereof being incapable of forming hydrogen bonds. If introduced in an aqueous environment, hydrophobic molecular units perturb or even disrupt the dynamic hydrogen bonded network that is formed by the water molecules. As a result, water molecules self-organize into more strongly ordered structures in the vicinity of the hydrophobic molecular units. This response is entropically unfavorable, and, hence, hydrophobic solutes become sticky and tend to cluster together. The solvent-mediated interactions at play are known as hydrophobic interactions (HIs). A distinguishing hallmark of HIs is their non-trivial dependence on temperature. For small hydrophobic solutes as well as large ones characterized by small hydrophobic patches, below, say, 1nm${}^{2}$, HIs become stronger with increasing temperature \cite{Chandler2005}. This characteristic feature of HIs persists in liquid water and distinguishes HIs from all other types of non-covalent attractive interactions that become effectively weaker at higher temperature. \par

HIs play a central role in various phenomena in chemistry and biology, from the cleaning action of detergents and the production of micro-emulsions, to the \textit{in vivo} assembly of biological macromolecules into complex structures. HIs often facilitate proteins in attaining their functional form by supporting their native fold or by binding to partners \cite{Dyson2006,Sturtevant1977}. HIs have been also implicated in promoting the self-assembly of proteins into oligomeric species and amyloid fibrils, a process accompanying many disease conditions \cite{Buell2012}. Finally, HIs, in an intricate interplay with other types of non-covalent interactions, drive the self-assembly of virus coat proteins into virus capsids \cite{Kegel2004}. This phenomenon has inspired the field of bionanotechnology to create novel self-assembled biosynthetic structures \cite{Comellas-Aragones2007, Hernandez-Garcia2014}. Apart from being a characteristic feature of HIs, the unique temperature response makes this type of non-covalent interaction a suitable tool to manipulate the properties of materials that have exposed patches of hydrophobic surface. Nevertheless, controlling material properties through hydrophobic forces remains a challenge, arguably resulting from our limited understanding of HIs and the lack of design principles for the synthesis of tunable materials, the responsiveness of which is based on HIs. \par 

Here, we make use of the neuronal protein alpha-synuclein ($\alpha$S) that under appropriate conditions self-assembles into amyloid fibrils, and take it as a model system in which material properties can be controlled by harnessing HIs. This protein exhibits a complex phase behavior and, depending on the physico-chemical conditions, organizes into hierarchical suprafibrillar aggregates with varying morphologies or into isotropic semi-flexible amyloid networks \cite{Semerdzhiev2014,Semerdzhiev2017}. We carefully choose the experimental conditions to steer the self-assembly into the region of the phase diagram where semi-flexible networks are formed. Fibril networks are a convenient platform i) to convincingly address the hydrophobic nature of the attractive interactions that drive the self-organization of $\alpha$S fibrils into larger scale structures, helping us to identify the role and formation mechanism of pathological fibril structures such as Lewy bodies that accompany the progression of Parkinson's disease, and ii) to exploit hydrophobic interactions to tune the mechanical properties of a material and inspire design principles for the creation of novel temperature responsive materials.
 We use temperature as a `tuning knob' to adjust the effective degree of crosslinking in the $\alpha$S fibril network and by doing so change the viscoelastic response of the material without forcing any permanent structural alterations in the material. Finally, we quantitatively connect the thermally induced enhancement of HIs to the observed stiffening in the viscoelastic response of the network by incorporating the effect of reversible crosslinking in established scaling theory for semi-flexible networks.\par

\section{Results and discussion}
\begin{figure}
\begin{center}
\includegraphics[width=86mm]{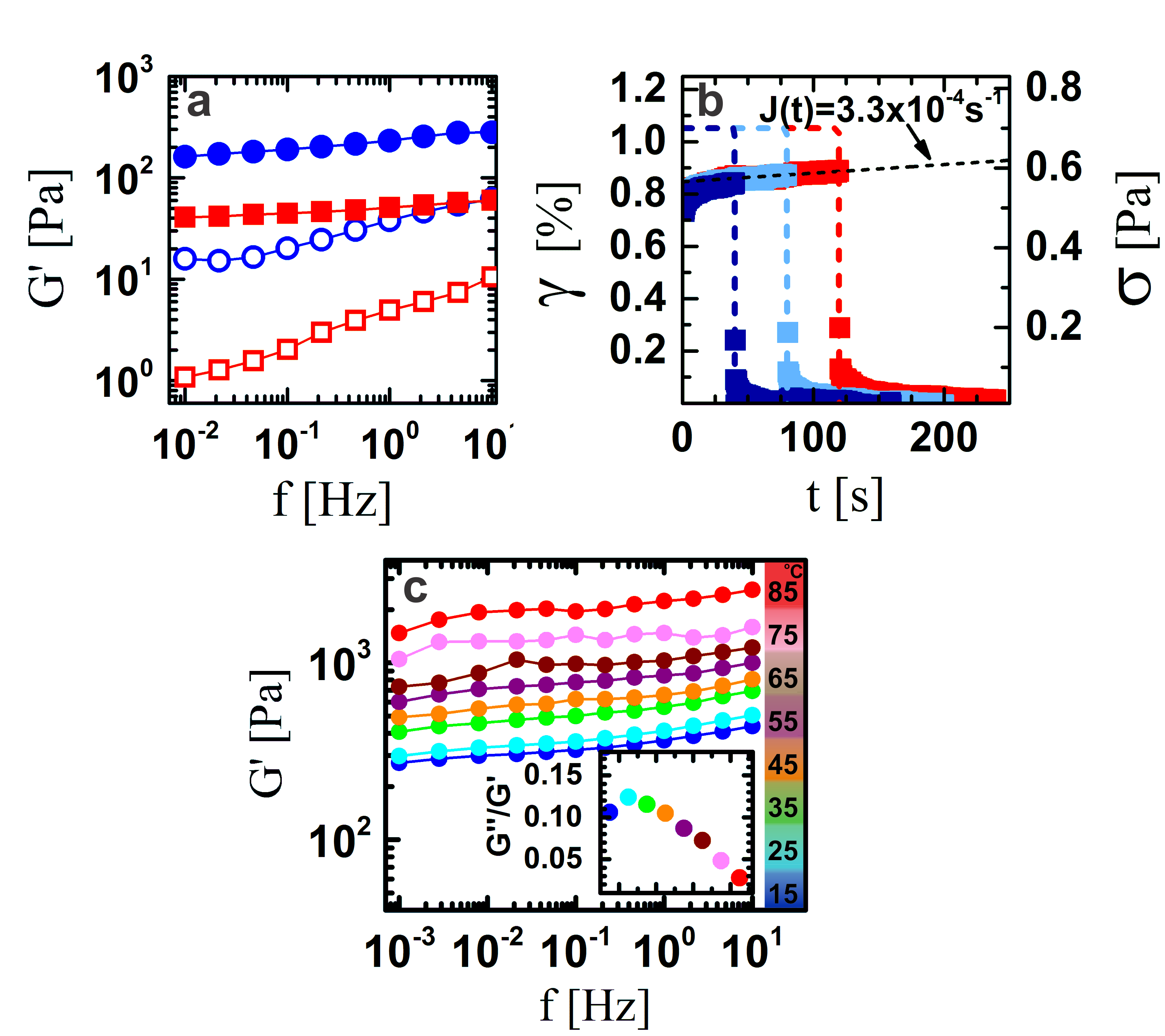}
\caption{\label{Fig. 1} Rheology on $\alpha$S amyloid networks. (a) Frequency sweeps for a 7 day (\raisebox{-.5ex}{$\textbf{\textcolor{red}\FilledSmallSquare}$}) and 37 days  (\raisebox{-.5ex}{$\textbf{\textcolor{blue}\FilledSmallCircle}$}) old sample. The storage modulus  and the loss modulus   are designated with closed and open symbols respectively. (b) Creep-recovery tests for an equilibrated 300 $\mu$M $\alpha$S amyloid network. Squares designate the measured strain  and the dashed block pulses represent the loading stages characterized by their duration and the amount of stress   (0.7 Pa) applied. Squares and dashed lines with different colors are used to discern the measured strain and the duration of the applied stress respectively for the 3 subsequent creep-recovery tests with increasing duration of the loading stage. The same sample was used for the three measurements. The estimate for the creep-compliance is obtained from the slope of the dashed line.(c) Frequency sweeps for an equilibrated network subjected to an extended temperature treatment. (inset) The damping factor $G'/G''$(tan$\beta$) at 1 Hz as a function of temperature. The decreasing value of the damping factor shows an increase in the elastic portion of the mechanical response of the network.}
\end{center}
\end{figure}

The polymerization of $\alpha$S into amyloid fibrils is a slow process. Within 7 days after the initiation of polymerization, a solution of monomeric $\alpha$S typically evolves into a gel (Fig. S1). It takes up to 37 days until all the protein has polymerized into fibrils (Fig. S2) and the network has equilibrated (Fig. \ref{Fig. 1}a). Rheologically, the networks of semi-flexible amyloid fibrils behave as viscoelastic materials. Frequency sweeps of aged networks produce relatively featureless spectra. No cross-overs between the frequency dependent storage modulus $G'(f)$  and loss modulus $G''(f)$ are observed in the probed frequency range (Fig.~\ref{Fig. 1}a). The amyloid network properties are dominated by the storage modulus, as is characteristic for viscoelastic solids. Both $G'$ and $G''$ are weakly dependent on the frequency, a feature typical of cross-linked polymeric materials. However, since the $\alpha$S amyloid networks are not chemically cross-linked, this observation implies significant attractive inter-fibril interactions, physically cross-link the fibrils. The presence of associative inter-fibril interactions is further supported by creep-recovery experiments (Fig.~\ref{Fig. 1}b). Applying a constant stress on an $\alpha$S fibril network initially induces a time-dependent strain response of the material. Shortly after the stress is applied, the change in the strain reaches a steady state known as creep. The $\alpha$S fibril gel exhibits very low creep (Fig.~\ref{Fig. 1}b), which is again in line with the presence of (localized) attractive inter-fibril interactions. Once the stress is removed, the network shows very low levels of plastic deformation and recovers almost completely to its original state. The same behavior is observed after extending the period during which the sample is subjected to stress (Fig.~\ref{Fig. 1}b).\par

Given the triblock copolymer-like architecture of the $\alpha$S monomer, consisting of an amphiphilic domain, a hydrophobic domain and a net charged domain, it is not surprising that this protein exhibits multiple modes of intermolecular interactions. While the stability of the amyloid fibrils is provided by hydrogen bonding, the driving force for the self-assembly into amyloid fibrils is believed to be hydrophobic interactions. The latter most probably also play a role in inter-fibril interactions \cite{Buell2012}. Considering that the $\alpha$S amyloid fold results from the subtle interplay between electrostatic interactions between the charged domains, hydrogen bonding and HIs, the minimum free energy conformation of the protein in the fibrils does not preclude some residual exposure of hydrophobic domains that can mediate HIs between fibrils \cite{Celej2008}. The formation of $\alpha$S fibril clusters after a high-temperature treatment of fibril suspensions indicates that HIs are indeed also involved at the inter-fibril level \cite{Semerdzhiev2014}. Plausibly, HIs are also responsible for the observed viscoelastic behavior of $\alpha$S networks (Fig. \ref{Fig. 1}a). \par

In the right settings, HIs can be made stronger by elevating the temperature of the system \cite{Chandler2005}. If HIs are indeed responsible for inter-fibril interactions in $\alpha$S amyloid networks, temperature should have a pronounced effect on the viscoelastic response of the network to applied deformations. Indeed, an $\alpha$S network significantly stiffens in the temperature range from 15 \textsuperscript{o}C to 85 \textsuperscript{o}C, which expresses itself in an order of magnitude increase of the storage modulus $G'$  (Fig. \ref{Fig. 1}c). This temperature-induced network stiffening is reversible. The value of  $G'$ tightly follows the changes in the temperature even if the network is repeatedly subjected to temperature cycles with different amplitudes (Fig. S3). This behavior is typically not observed in networks of semi-flexible polymers, yet does superficially resemble the elastic behavior of rubbers. In the latter, the free energy cost of stretching out the stored contour length of the crosslinked polymers becomes much larger at elevated temperatures, an effect associated with them being flexible rather than semi-flexible. However, $\alpha$S fibrils formed at the conditions used to prepare the amyloid gels appear to be very stiff as is evidenced by TIRF microscopy images (Fig. S4a).  An end-to-end distance versus contour length analysis of the fibrils yields an estimate for the persistence length of  $l_p\approx$ 85 $\mu$m, which is very much larger than the mesh size expected for a 300 $\mu$M $\alpha$S fibril network (Fig. S4b and S4c). It is therefore unlikely that the increased free energy cost of reducing the conformational freedom of an individual fibril contributes significantly to the observed increase in $G'$  with increasing temperature. \par

\begin{figure*}
\includegraphics[width=100mm]{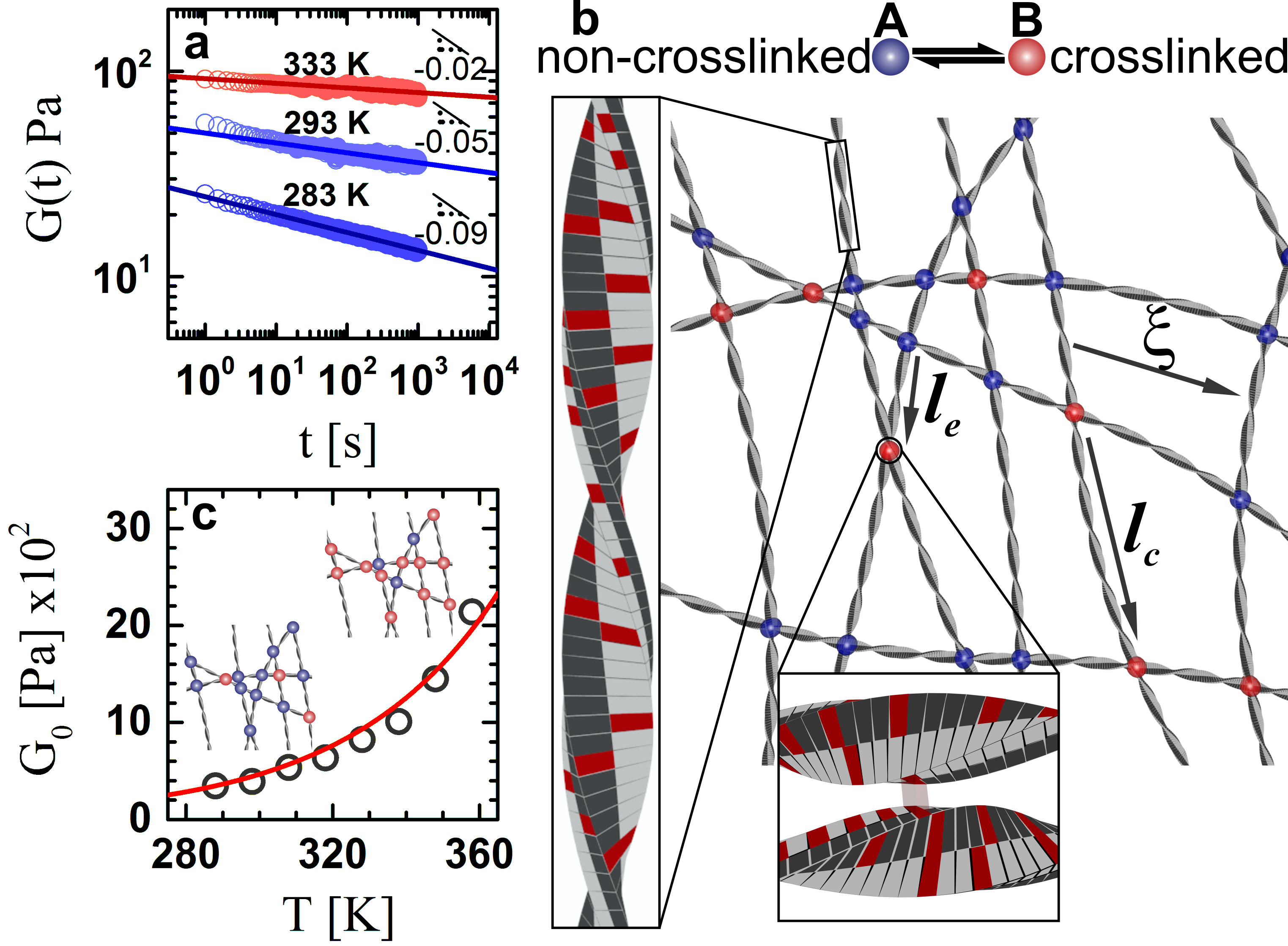}
\caption{\label{Fig. 2} Hydrophobically crosslinked $\alpha$S amyloid networks. (a) Stress relaxation tests of a 300 $\mu$M $\alpha$S network (2 mM Na\textsuperscript{+}) at different temperatures. The stress-relaxation curves are vertically shifted for a better visualization. (B) Artist impression of a ‘hydrophobically crosslinked’ $\alpha$S amyloid network. The tentative hydrophobic patches on the surface of the fibrils are presented in red. (c) Scaling of the storage modulus with temperature. The circles represent the experimental data, the red curve is the fit generated using the scaling relation derived in the main text (eq. \ref{eq. 2} ). (cartoon insets) At higher temperature the number of effective cross-links is significantly higher as compared to lower temperatures.}
\end{figure*}

A huge experimental and theoretical research effort has been invested in the past few decades to better understand the viscoelastic properties of networks comprised of semi-flexible polymers \cite{Broedersz2014}. These efforts have resulted in theoretical frameworks that describe scaling relations between quantities characterizing the properties of these networks, and which have been successfully applied to a wide range of biological and synthetic materials, all falling in the general class of semi-flexible polymer networks \cite{Gardel2004,Huisman2011,Kouwer2013}. In view of the featureless frequency sweeps and the creep recovery experiments, the  $\alpha$S fibril network seems to behave as a crosslinked network (Fig. \ref{Fig. 1}a). With this in mind, we have adopted the scaling theory for cross-linked semi-flexible networks in order to interpret the temperature behavior of the $\alpha$S amyloid network. Because of the relatively large persistence length of $\alpha$S fibrils compared to the mesh size, we invoke a scaling theory based on the so-called floppy modes model, assuming a constant strain \cite{Heussinger2006}. According to this model, we have:

\begin{equation}
\label{eq. 1}
G'_0=\frac{\kappa}{\xi^2 l_c^4}
\end{equation}

where $G'_0$   is the plateau modulus of the network, $\kappa$ is the bending stiffness of the semi-flexible chains, $\xi$ the average mesh size and $l_c$ the average distance between crosslinks \cite{Lieleg2007} (see also Fig. \ref{Fig. 2}b). Note that the persistence length and bending stiffness are related according to $l_p=\kappa/k_B T$. The strong dependence of the plateau modulus on the number of crosslinks (through $l_c$) is apparent. However, before focusing on this particular quantity, the possible contribution of the other two relevant parameters to the observed thermal stiffening in $\alpha$S networks, namely $\xi$ and $\kappa$  , needs to be considered. Large temperature-induced stiffening of semi-flexible polymers has been observed in some synthetic systems. Driven by the enhanced hydrophobic interactions at higher temperatures, synthetic polymers may bundle into filaments with more than an order of magnitude larger rigidity \cite{Kouwer2013}. The enhancement of  $\kappa$ can in that case be large enough to overcome the effect of the increased mesh size, which is expected to soften the network (eq. \ref{eq. 1}) and to produce an overall increase in the plateau modulus \cite{Tharmann2006}. Even though hydrophobic interactions also seem to play an important role at the intra- and inter-fibril level in $\alpha$S networks, bundling is an unlikely mechanism to account for the experimental observations at higher temperatures. SAXS measurements do not provide any evidence for significant structural changes in the $\alpha$S network at these higher temperatures. The size of the fibril’s cross-section remains close to constant throughout the temperature ramps (Fig. S4d and S4e). Moreover, the SAXS curves remain identical at the different temperatures indicating that there are no sizable changes in the overall structure of the network and consequently in the mesh size $\xi$.\par

If bundling does not take place, then the strong temperature dependence of  $G'_0$ might result from a drastic change in the bending rigidity of the individual fibrils themselves with temperature. Since $G'_0\sim\kappa$  (eq. \ref{eq. 1}), the observed change in $G'_0$  between 5 - 80 \textsuperscript{o}C would imply a 13-fold increase over that temperature range. This estimate is based solely on changes in $\kappa$ without taking into account the changes in the so-called entanglement length $l_e$. If the sensitivity of  $l_e$ on the temperature is also taken into account, using the known scaling relations and assuming crosslinks can only appear at entanglement points $l_c\sim l_e\sim l_p\sim (\kappa/k_bT)$ , then the increase of $\kappa$  with temperature would need to be even larger (eq. \ref{eq. 1}) \cite{Isambert1996}. The temperature dependence of $\kappa$ is however generally very moderate for semi-flexible biopolymers. Additionally the sign of the change strongly depends on the biopolymer species. While, depending on GC content and salt concentration, double-stranded DNA seems to exhibit a 15-20 $\%$ reduction in $\kappa$ with increasing temperature by 30-45 \textsuperscript{o}C, single-stranded DNA shows a 15$\%$ increase of $\kappa$ over a comparable temperature range \cite{Lorenzo2015,Geggier2011,Driessen2014}. A drastic change in the mechanical properties of individual $\alpha$S fibrils also seems highly unlikely. As mentioned earlier, the protein monomers in the cross $\beta$-sheets fibril backbone are held together by numerous intermolecular hydrogen bonds, which are the main determinant for the fibril stiffness \cite{Fitzpatrick2015}. Driven by inter- and intra-molecular HIs, $\alpha$S monomers aggregate and attain a fold with optimized internalization of apolar residues in the fibril’s core. Keeping this in mind, higher temperatures stimulate two counteracting effects: on one hand, the breaking of hydrogen bonds should reduce the rigidity of the fibrils and, on the other, the enhancement of the hydrophobic interaction could increase the stiffness. However, it is unlikely that the strength of inter-monomer HIs increases sufficiently to compensate for the loss of hydrogen bonds, and produce the dramatic net increase in $\kappa$  that would account for the unusual increase in the $G'_0$  of the network. Indeed, other molecular assemblies that are held together by hydrophobic interactions also do not show signs of unusual stiffening, induced by an increase in temperature in the range comparable to the one used in this study. Lipid bilayers, for example, become easier to deform with increasing temperature  \cite{Niggemann1995,Pan2008}. The filamentous fd virus exhibits a non-monotonic change in the persistence length $l_p=\kappa/k_bT$ with temperature: at higher temperature, $l_p$ decreases, while an increase is found at the lower temperature range \cite{Tang1996}. This variation in $l_p$  is however small, amounting to no more than 30$\%$.\par

With changes in $\kappa$  being an unlikely cause for the large temperature-induced increment of $G'_0$, the only parameter left that could potentially account for this phenomenon is the mean distance between crosslinks $l_c$. Heating up the system seems to strengthen the hydrophobic contacts between the fibrils, which ultimately results in a more densely crosslinked network. Results from stress-relaxation measurements on an $\alpha$S network at different temperatures are in line with this hypothesis. Instead of speeding up the relaxation processes, heating up the sample actually slows down the relaxation dynamics (Fig. \ref{Fig. 2}a), probably due to the enhanced inter-fibrillar contacts (Fig. \ref{Fig. 2}b).\par
To test the hypothesis that heating the amyloid network strengthens hydrophobic contacts between fibrils, we establish a quantitative relation between temperature and the storage modulus. For this purpose, the impact of temperature on the effective number of crosslinks in the system is evaluated at the level of a Boltzmann equilibrium and incorporated in the scaling relations for crosslinked semi-flexible networks (for the derivation see the Supplemental Material). This results in the following relation: 

\begin{equation}
\label{eq. 2}
G'_0(T)=G'_0(T_0)e^{\frac{H_0}{k_bT_0^2}(T-T_0)}\Bigg(\frac{T_0}{T}\Bigg)^{(2/5)}
\end{equation}

where $G'_0(T)$ is the temperature-dependent plateau modulus, and  $G'_0(T_0)$ the plateau modulus at the reference temperature $T_0$ . 
The value of  $H_0$ strongly depends on the architecture of $\alpha$S fibrils. Since there is no established model for this architecture, $H_0$  is left as a free parameter. The reference temperature $T_0=288\ K$ , which is the lowest temperature at which the storage modulus was measured, is used to fit equation \ref{eq. 2} to the experimental data. Equation \ref{eq. 2} seems to describe the experimental observations very well indeed (Fig. \ref{Fig. 2}c). The fit yields an endothermic value for $H_0=7.5\  k_bT$. From the obtained value for $H_0$  we can estimate the apparent size of the hydrophobic patches using the expression for the enthalpy at the reference state of hydrophobic contacts: $H_0=2h_{h.i.}a$ where $h_{h.i.}$  is the energy cost per unit area  of exposed hydrophobic surface and $a$ is the area \cite{Phillips2009}. Taking into account that typically $h_{h.i.}\sim 7 \ k_bT nm^{-2}$  , the estimate for the size of the hydrophobic patches on the fibril surface   is $\sim 0.75 \ nm^2$ which is comparable to what has been found previously for virus coat proteins \cite{Kegel2004,Phillips2009,Kraft2012}.\par

\section{Conclusion}
In summary, $\alpha$S amyloid networks exhibit remarkable thermo-responsive properties. The fibrillar gel significantly stiffens at higher temperatures and completely recovers its original state once the temperature is lowered again. We propose that the thermo-stiffening of the $\alpha$S network is the consequence of enhanced inter-fibrillar hydrophobic contacts stimulated by the higher temperature. This is consistent with previously established qualitative findings, suggesting that the hydrophobic effect plays an essential role in the interaction between $\alpha$S fibrils \cite{Semerdzhiev2014}. The presence of hydrophobic interactions between fibrils suggest that multiple hydrophobic domains in the fibril core remain solvent exposed. At higher temperatures these hydrophobic areas become “activated”, which effectively increases the number of contacts points between fibrils. 
An alternative explanation in which the exposure of the hydrophobic domains itself is a temperature-induced phenomenon could also be considered.  At higher temperature segments of the $\alpha$S fibrils may unfold and reveal the hydrophobic stretches of the protein sequence to the solvent, which later become anchoring points between fibrils. Such a hypothetical scenario would be consistent with previous research, suggesting that at elevated temperatures the cross-beta sheet structure of the fibrils starts to fall apart \cite{Ikenoue2014}. However, the experimental findings reported here do not give any clues supporting this interpretation. Numerous unfolding events in fibrils should have become apparent in the SAXS cross-sectional Guinier analysis, as it changes the effective cross-section of the fibril. We do not observe this in our SAXS data. Moreover, compromising the structural integrity of the fibrils should also have altered the mechanical response of the network towards softening rather than towards stiffening.Elucidating the cohesive forces between amyloid fibrils is crucial for obtaining a better understanding of the associated pathology and the physiological role of such structures. A correlation between the exposure of hydrophobic surface in amyloid aggregates and their toxicity has been suggested by numerous studies \cite{Bemporad2012,Mannini2014,Olzscha2011}. Considering the nanoscale organization of the $\alpha$S fibrils it is unlikely that all hydrophobic patches on the fibril surface are protected from contact with the aqueous environment by inter-fibril interactions. The exposure of these hydrophobic fibril patches to the cytosol may induce interactions with other proteins. The accumulation of additional proteins in amyloid deposits may therefore not be a result of preserved functional interactions but rather be an effect of HIs. The accumulation of amyloid fibrils might increase the total hydrophobic surface present in the cell and thereby interfere with its normal functioning.\par
Understanding the inter-fibril interactions is also imperative for the successful utilization and manipulation of amyloid materials. Our data indicate that there are opportunities to harness these interactions and tune the mechanical properties of amyloid materials. Moreover, these findings indicate that it should be possible to design amyloid fibrils or other supramolecular assemblies with engineered hydrophobic patches and synthesize materials with imprinted temperature responsiveness. An important question remains, however. Are these interactions generic for amyloids or just specific for $\alpha$S? Exposure of amyloid networks comprised of the disease-unrelated protein $\beta$-lactoglobulin does not seem to provoke the same response, indicating that the degree of thermo-stiffening observed for $\alpha$S gels cannot be expected to hold for all amyloid materials \cite{Gosal2004}. 

\section{Acknowledgements}
This work was supported by the “Nederlandse Organisatie voor Wetenschappelijk Onderzoek” (NWO) through NWO-CW Veni grant (722.013.013) to S.L., NWO-CW TOP program (700.58.302) to V.S., NWO-CW VIDI grant (700.59.423) to M.M.A.E.C., and through and Nanonext NL theme 8A. The authors thank Kirsten van Leijenhorst-Groener and Nathalie Schilderink for the expression and purification of α- synuclein. SAXS experiments were performed at ESRF, BM26B (DUBBLE) experiment number 26-02-664. We thank Dr. G. Portale (Beamline Scientist) for his help performing SAXS experiments.

\newpage

\onecolumngrid

\section{Supplemental Material}
\subsection{I. Materials and methods}
\subsubsection{A. $\alpha$S gel preparation}
Expression of the human wild type $\alpha$S was performed in \textit{E. coli} B121 (DE3) using
the pT7-7 based expression system. Details on the purification procedure for $\alpha$S
are described elsewhere \cite{Rooijen2009b}. Gels of $\alpha$S amyloid fibrils were prepared in
quiescent conditions. Fibril growth was seeded by 5 mol.$\%$ preformed $\alpha$S seeds at
a total protein concentration of 300 $\mu$M $\alpha$S in 10 mM Tris, pH 7.4 and 10 mM NaCl
(Sigma). The first 6 days samples were incubated at 37 \textsuperscript{o}C and after that the
gels were stored to mature at room temperature. The gels for SAXS measurement
were directly grown in quartz capillaries (Hilgenberg GmbH, L=80, OD=1.5,
Wall=0.01 mm) using the procedure described above. Samples for the TIRFM imaging
were stained with the amyloid specific fluorescent dye Thioflavin T (ThT) and
sealed in custom made glass chambers.

\subsubsection{B. $\alpha$S gel preparation}
A solution of fibrils was prepared by incubating 100 $\mu$M of $\alpha$S in 10
mM Tris (Sigma), 2 mM NaCl (Sigma), pH=7.4, 37 \textsuperscript{o}C and shaking
at 900 RPM. Once the aggregation was completed, the fibril solution was
sonicated (Branson, Sonifier 250) on ice. The solution was sonicated at the
lowest power for 5 seconds and left at rest for another 55 second. The cycle was
repeated five times. Subsequently, the sonicated samples were tested for seeding
efficiency by incubation with $\alpha$S monomers.

\subsubsection{C. SAXS}
Experiments were performed at the BM26 DUBBLE (Dutch-Belgian Beamline, ESRF,
Grenoble, France). Two dimensional images were collected using Pilatus 1M photon
counting detector. The sample to detector distance was 6.6 m. The wavelength for
the incident x-ray was 0.1 nm\textsuperscript{-1} and beam cross-section with
dimensions 2.5 mm x 4.5 mm. The energy of the x-rays was 12 eV. The attained $q$
range was 0.03−1.5 nm\textsuperscript{-1}. The ATSAS 2.6.0 software package was
used for post-acquisition processing and analysis of the SAXS data
\cite{Petoukhov2012}.

\subsubsection{D. Rheology}
Rheology measurements were performed on an Anton Paar MCR 301 rheometer using a
plate-plate geometry, with a plate diameter of 25 mm. Gel samples were carefully
collected from the storage tubes using truncated pipette tips to minimize
shearing and network disruption. The gel was then carefully deposited on the
rheometer’s stage. Measurements were conducted at gap size of 0.16 mm and the
samples were covered with mineral oil to avoid evaporation during the
temperature runs. For each temperature step the sample was first left to
equilibrated until the storage modulus $G'$ measured at $f= \ $0.5 Hz, $\gamma=\
$0.5$\ \%$, attained a stationary value (usually within 1 hour). Subsequently, a
frequency spectrum was recorded at $\gamma=\ $1$\ \%$. Stress relaxation test
were conducted by first equilibrating the sample at a given temperature and then
subjecting it to a strain $\gamma=\ $3$\ \%$ which is within the linear
viscoelastic response region of the sample. Subsequently the time relaxation
modulus $G(t)$ was recorded.

\subsubsection{E. Total Internal Reflection Fluorescence Microscopy (TIRFM)}
The TIRFM imaging was performed using a Nikon Ti-E microscopy setup coupled with
an Argon laser (35-IMA-040, Melles Griot, USA). Images were acquired with a CFI
Apo TIRF 100x objective (Nikon, Japan) and iXon 3 DU-897 EMCCD camera (Andor,
UK) using the 457 nm line of the laser for the excitation of the ThT dye. The
used filter cubes contained a 455 nm excitation filter with 10 nm bandpass, a
458 nm long pass dichroic mirror and a 485 nm emission filter with 30 nm bandpas

\subsubsection{F. Determination of residual monomer concentration}
Determination of residual monomer concentration. Samples of the S amyloid gels
were centrifuged using a Sorvall WX 80 ultracentrifuge (Thermo Scientific, USA)
and Fiberlite F50L-24 x 1.5 fixed-angle rotor at 247 kG for 5 hours.
Subsequently the concentration of free $\alpha$S monomers in the supernatant was
determined using a UV-Vis spectrophotometer (UV 2401 PC, Shimadzu). Absorbance
was measured at 276 nm and an extinction coefficient of 5600
M\textsuperscript{-1}.cm\textsuperscript{-1} was used to calculate the protein
concentration.

\subsubsection{G. Persistence length analysis}
Aliquots of a 300 $\mu$M $\alpha$S gel, 10 mM NaCl and 10mM Tris were diluted in
the same buffer keeping the ionic strength constant. Thioflavin T (ThT) was
subsequently added to a final concentration of 1 $\mu$M in order to stain the
fibrils. A small volume of the diluted sample was sealed between a glass slide
and a cover slip. The sample was left at rest for the fibrils to sediment on the
cover slip and then imaged using Nikon Ti-E in TIRF mode (see section TIRFM).
Images were analyzed using the MATLAB based package Easyworm \cite{Lamour2013}.
Briefly, the persistence length is determined via random resampling using
bootstrap with replacement method. A bootstrap sample of n chains is randomly
selected from the ensemble of all chains. For each bootstrap the average square
of the end to end distance $<R^2>$ is binned at equal length intervals. Then all
the data is fitted using the wormlike chain model:

\setcounter{equation}{0}

\begin{equation}
\renewcommand{\theequation}{S\arabic{equation}}
\label{eqS1}
<R^2>=2l_pL\Bigg[1-\frac{l_p}{L}\bigg(1-e^{\frac{-L}{l_p}}\bigg)\Bigg]
\end{equation}

where $L$ is the contour length.

\subsubsection{H. Derivation of the equation linking hydrophobic interactions to the
temperature induced stiffening of $\alpha$S fibril networks}
Temperature seems to enhance the formation of physical crosslinks between
fibrils in the $\alpha$S amyloid network. Identifying the impact of temperature
on the effective number of crosslinks in the system, would enable us to
incorporate the temperature effect in the scaling relations for the elastic
response of crosslinked semi - flexible networks. To directly relate the
enhanced hydrophobic interactions to the stiffening of the network, we introduce
a two state model for the entanglements in the network $A\rightarrow B$, where
$A$ and $B$ represent the `free' and `crosslinked' entanglements respectively
(Fig 4B from the main article). Assuming that hydrophobic interactions drive the
crosslinking of entanglements, the equilibrium dissociation constant can be
written as \cite{Kegel2004,Schoot2005}:

\begin{equation}
\renewcommand{\theequation}{S\arabic{equation}}
\label{eqS2}
K=K_{T_0}e^{\frac{H_0}{k_bT_0^2}(T-T_0)}
\end{equation}

where $K_{T_0}$ is the equilibrium dissociation constant at a reference
temperature $T_0$, and $H_0>0$ is the endothermic binding enthalpy at
temperature $T_0$ for two hydrophobic surfaces coming into contact. Because
$H_0>0$ , $K$ increases with $T$ as is typically observed for hydrophobic
surfaces \cite{Kegel2004}. We can calculate the fraction $F$ of entanglements
that are in the crosslinked state by invoking Boltzmann statistics:

\begin{equation}
\label{eqS3}
\renewcommand{\theequation}{S\arabic{equation}}
F=\frac{K_{T_0}e^{\frac{H_0}{k_bT_0^2}(T-T_0)}}{1+K_{T_0}e^{\frac{H_0}{k_bT_0^2}(T-T_0)}}
\end{equation}

Since $K_{T_0}$ is presumably small, at low $T_0$ the hydrophobic interactions
are known to be weak \cite{Dapoian1995} so is $|T-T_0|/T_0$ over the probed
temperature range, we can approximate $F$ making use of a Taylor expansion:

\begin{equation}
\label{eqS4}
\renewcommand{\theequation}{S\arabic{equation}}
F\approx K_{T_0}e^{\frac{H_0}{k_bT_0^2}(T-T_0)}
\end{equation}

In a semiflexible network, contacts between filaments appear at the entanglement
points. Given that i) there are no significant structural rearrangements in the
network during the crosslinking process, ii) the maximum number bonds equals the
number of entanglement points, we conclude that the distance between crosslinks
$l_c$ can be related to the entanglement length $l_e$ through the fraction $F$:

\begin{equation}
\label{eqS5}
\renewcommand{\theequation}{S\arabic{equation}}
l_c\sim \frac{l_e}{F}
\end{equation}

Taking into account that $\xi \sim c_p^{-2}$, where $c_p^{-2}$ represents the concentration of the filament
forming protein, the entanglement length reads as \cite{Degennes1976}:

\begin{equation}
\label{eqS6}
\renewcommand{\theequation}{S\arabic{equation}}
l_e \sim l_p^{1/5} \xi^{4/5} \sim \Big( \frac{\kappa}{k_bT} \Big)
\end{equation}

where $\kappa$ is the bending stiffness of the semi-flexible chains and $\xi$ the average mesh size. We know that for a network comprised of semi-flexible polymers that appear stiff between entanglement points, the storage modulus scales as (see eq. 1 and  the discussion preceding it the main article):

\begin{equation}
\label{eqS7}
\renewcommand{\theequation}{S\arabic{equation}}
G'_0=\frac{\kappa}{\xi^2 l_c^4}
\end{equation}

where $G'_0$   is the plateau modulus of the network. Substituting equations eq. S\ref{eqS4} and eq. S\ref{eqS6} in eq.S\ref{eqS5}, inserting the latter in eq. S\ref{eqS7}, and finally after some re-arrangements we arrive at: 

\begin{equation}
\label{eqS8}
\renewcommand{\theequation}{S\arabic{equation}}
 G'_0(T)=G'_0(T_0)e^{\frac{H_0}{k_bT_0^2}(T-T_0)}\Bigg(\frac{T_0}{T}\Bigg)^{(2/5)}
\end{equation}	 
\clearpage

\subsection{II. Figures}
\setcounter{figure}{0}  

\begin{figure*}[htb] \centering
\vspace{5cm}
\includegraphics[width=86mm]{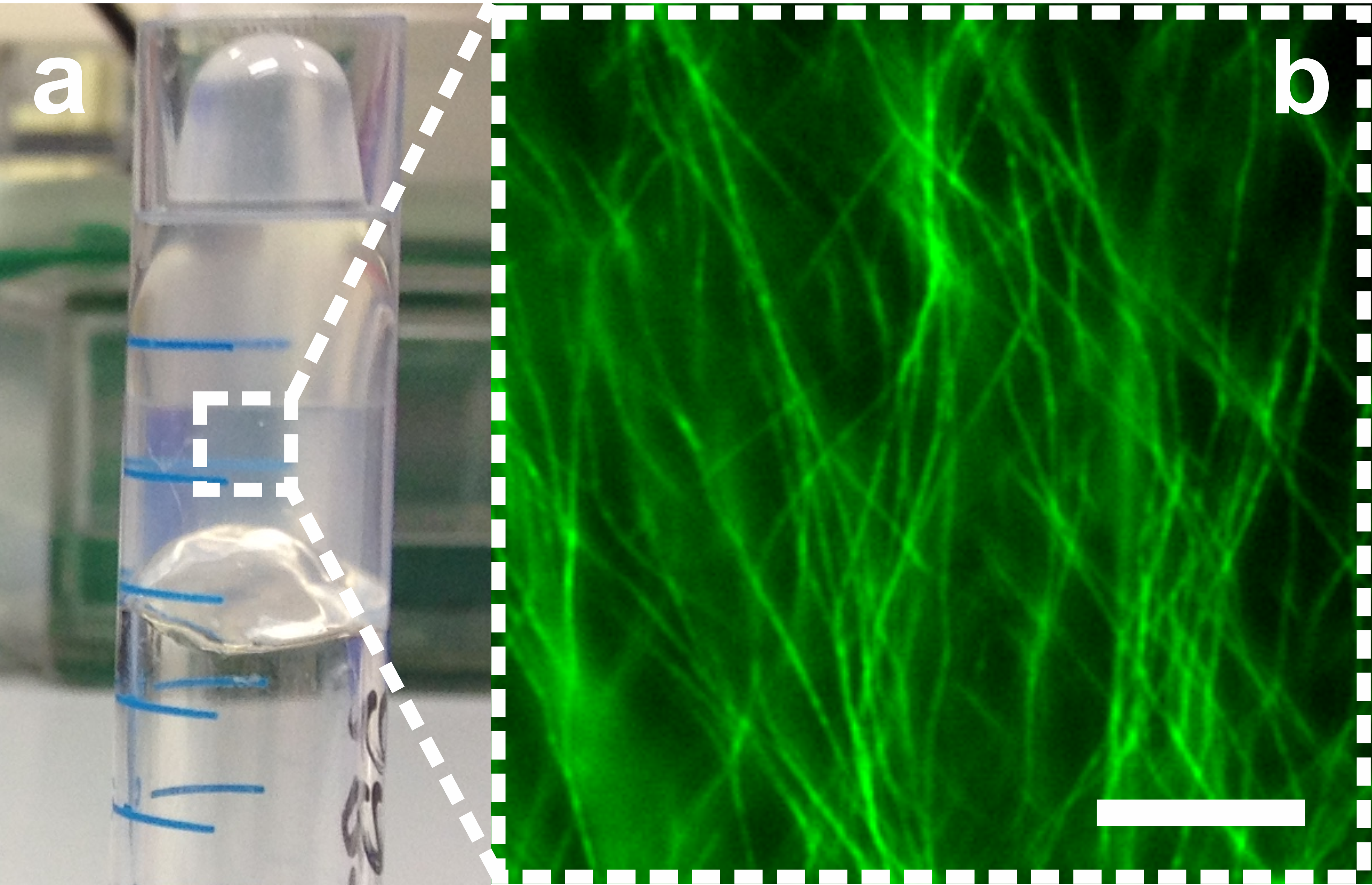}
\renewcommand{\thefigure}{S\arabic{figure}}
\caption{\label{Fig. S1} $\alpha S$ amyloid networks. (a) Self-supporting 300 $\mu$M $\alpha$S fibril network subjected to an inversion test. The gel was formed in 10 mM NaCl, 10 mM Tris and pH=7.4. (b) Total internal reflection microscopy (TIRF) images of an $\alpha$S network formed at the same conditions confirm that the gel is comprised of a network of amyloid fibrils.The network is stained with ThT (30 $\mu$M). The scale bar is 5 $\mu$m.}
\end{figure*}
\clearpage

\begin{figure*}[htb] 
\vspace{5cm}
\includegraphics[width=80mm]{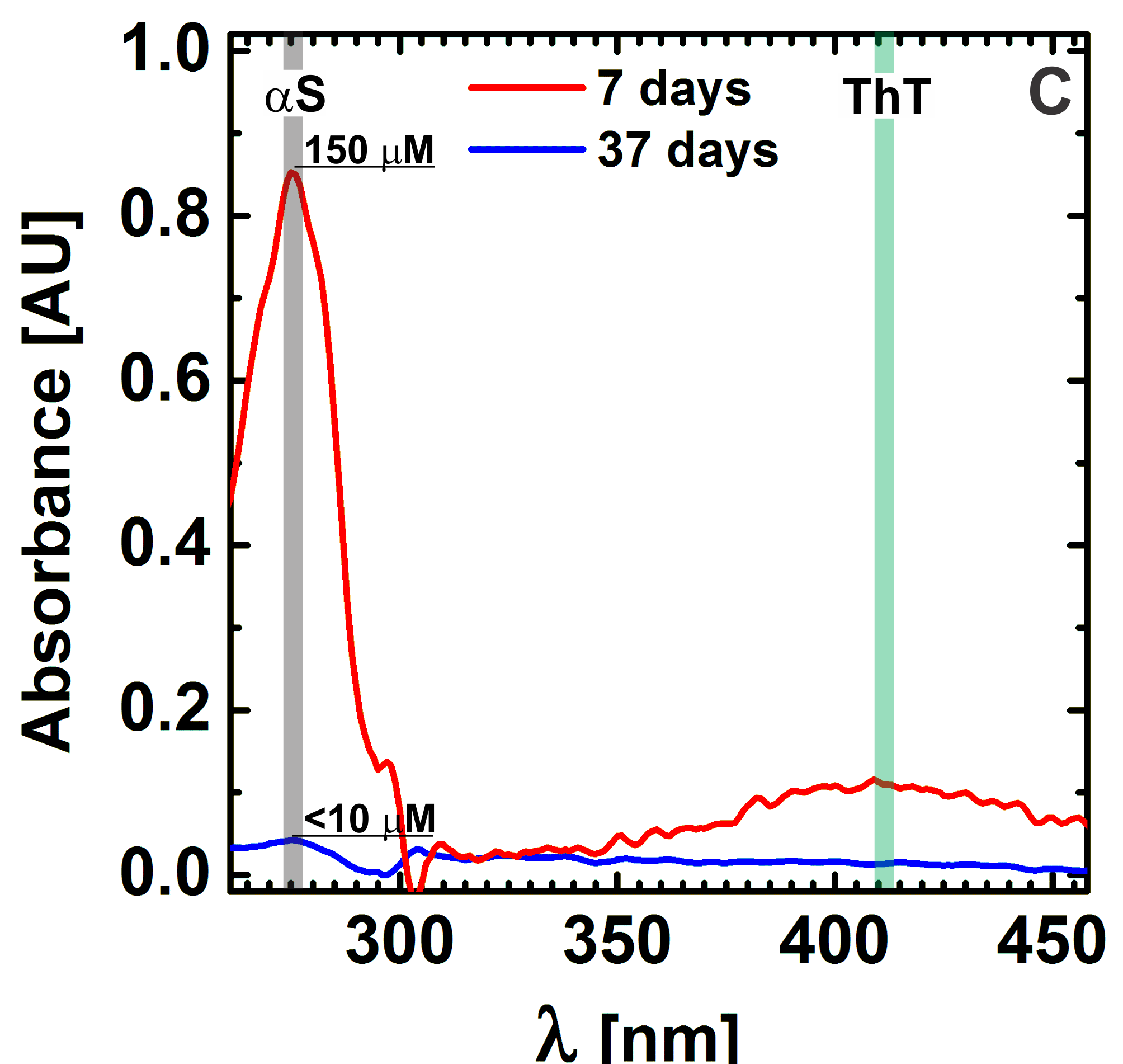}
\renewcommand{\thefigure}{S\arabic{figure}}
\caption{\label{Fig. S2} Absorbance spectra of the supernatant obtained after
the ultracentrifugation of an 300 $\mu$M $\alpha$S gel at different time points
of the aging process. After 7 days, only half of the monomers have been
recruited in fibrils and a detectable amount of free ThT is present in the
solution. At 37 days the residual level of monomeric protein is below the
threshold for an accurate determination and no free ThT is detected.}
\end{figure*}

\clearpage
\begin{figure*}[htb] \centering
\vspace{5cm}
 \includegraphics[width=70mm]{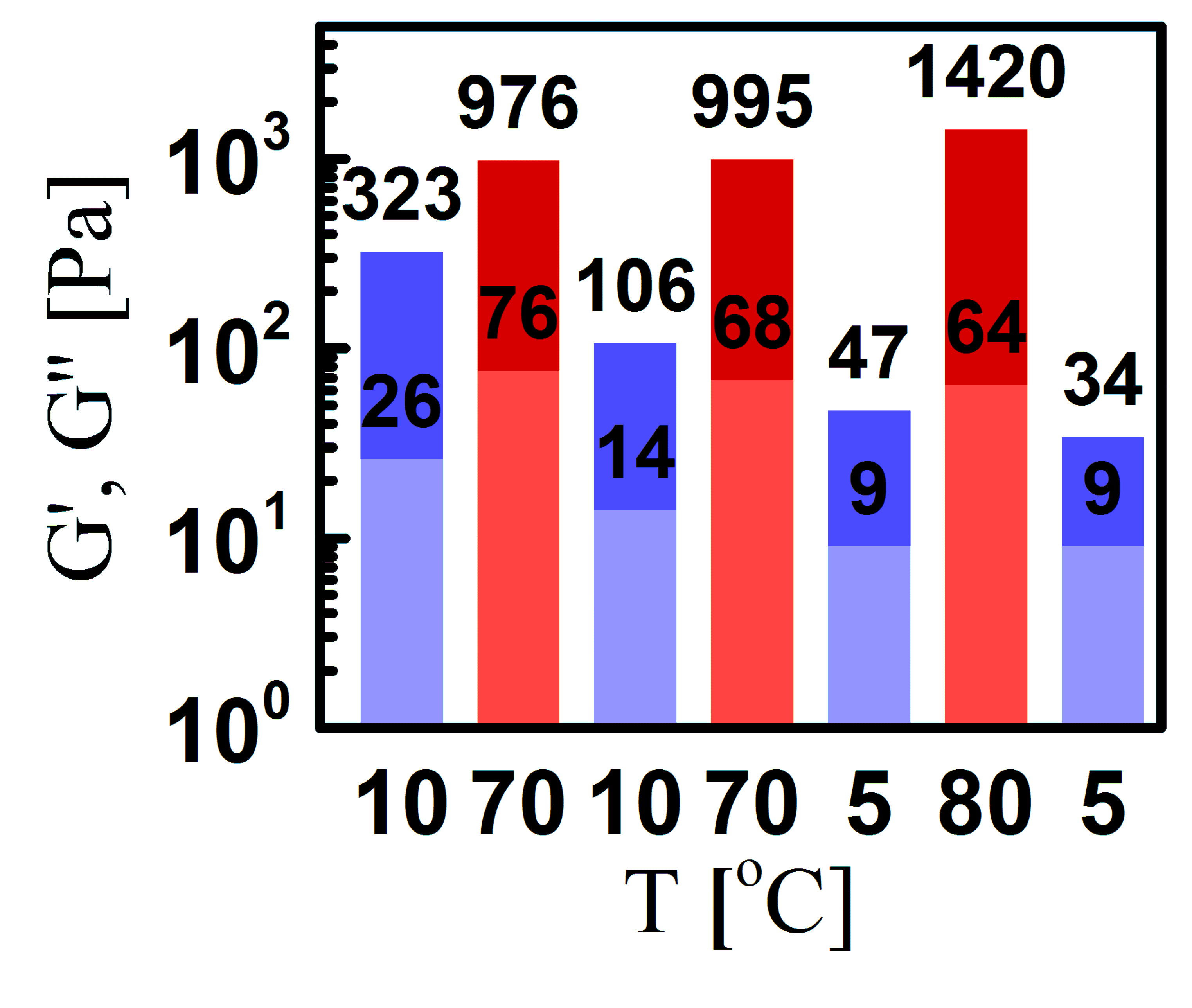}
\renewcommand{\thefigure}{S\arabic{figure}}
\caption{\label{Fig. S3} Reversible temperature stiffening of $\alpha$S amyloid networks. An equilibrated $\alpha$S amyloid network that is repeatedly subjected to cycles with different temperature amplitudes. Blue and red bars designate low and high temperature cycles respectively. Dark and light colored bars refer to $G'$  and $G''$  respectively .}
\end{figure*}

\begin{figure*}[htb] \centering
\vspace{2cm}
\includegraphics[width=110mm]{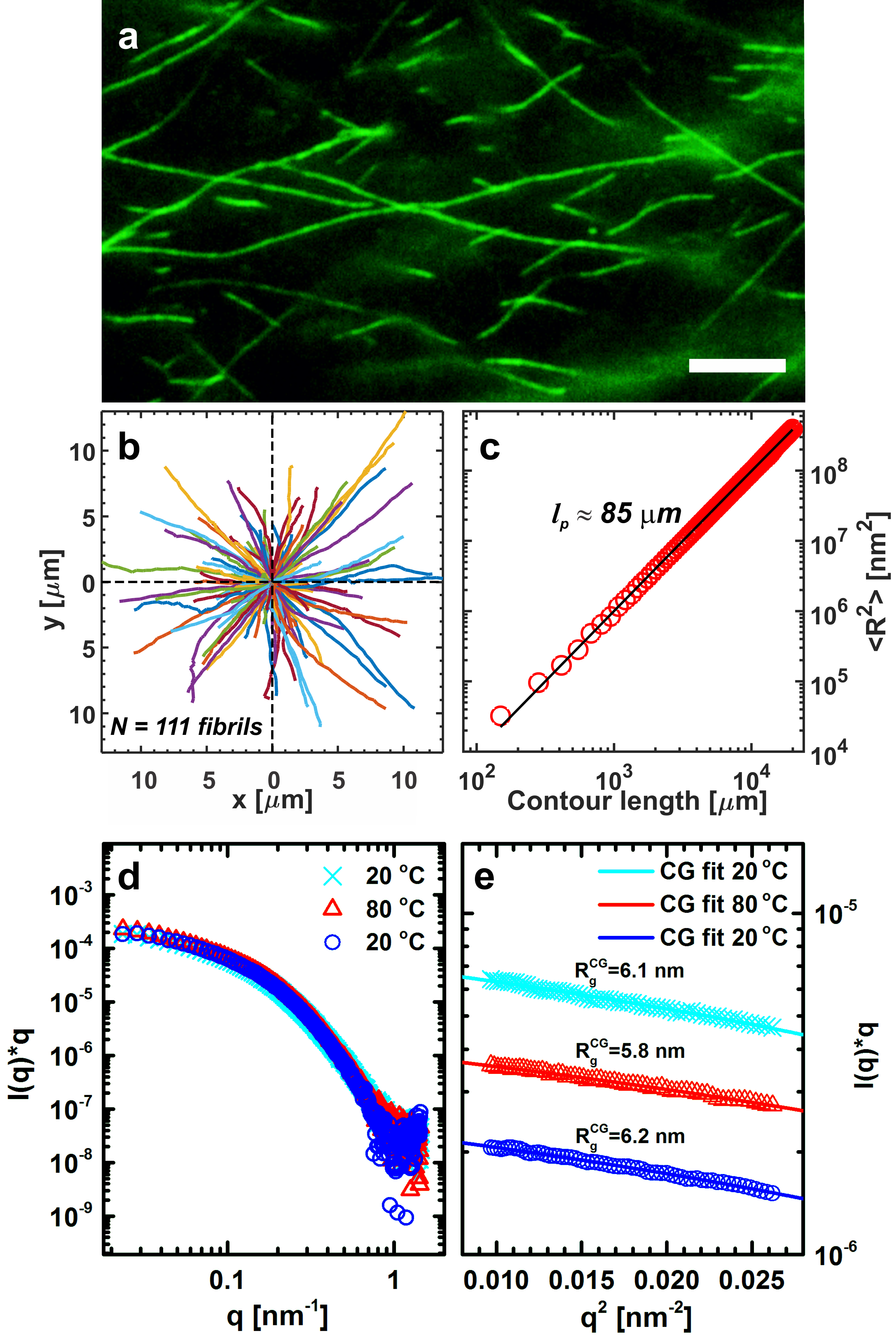}
\renewcommand{\thefigure}{S\arabic{figure}}
\caption{\label{Fig. S4} Mechanical and SAXS characterization of $\alpha$S fibrils. (a) TIRF image of individual $\alpha$S fibrils. Scale bar is 5 $\mu$m. (b) Traces for the set of analyzed fibrils. (c) Contour length vs mean square of the end to end distance. SAXS measurements of a network at different temperatures. (d) 1D SAXS curves for an 35 day old 300 $\mu$M $\alpha$S network. The sample was first measured at 20 \textsuperscript{o}C (\raisebox{-.5ex}{$\textbf{\textcolor{mycyan}\SmallCross}$}) and subsequently heated up to 80 \textsuperscript{o}C (\raisebox{-.5ex}{$\textbf{\textcolor{myred}\SmallTriangleUp}$}) and measured again.  The final measurement was performed after the sample was cooled down to the starting temperature (\raisebox{-.5ex}{$\textbf{\textcolor{myblue}\SmallCircle}$}). (e) Cross-sectional Guinier (CG) plots. The fibril thickness seems constant before, during and after the temperature treatment of the equilibrated $\alpha$S network. Data sets are vertically shifted for a better visualization.}
\end{figure*}

\clearpage
\section{References}

%

\end{document}